 \RequirePackage[2020-02-02]{latexrelease}
\documentclass[aps,prl,twocolumn,groupedaddress,showpacs,english]{revtex4}

\usepackage{placeins}
\usepackage[T1]{fontenc}
\usepackage[latin9]{inputenc}
\usepackage{amsmath}
\usepackage{amssymb}
\usepackage{graphicx}
\usepackage{subfig}
\usepackage{color,soul}

\usepackage{caption}
\usepackage{appendix}
\usepackage{wasysym}
\usepackage[version=4]{mhchem}
\usepackage{collcell}

\usepackage{comment}

\usepackage{babel}
\usepackage{chngcntr}
\begin{document}
\title{Purcell factors and Forster resonance energy transfer in proximity to  helical structures
}
\author{Asaf Farhi\textsuperscript{1,2} and Aristide Dogariu\textsuperscript{1}}
\affiliation{\textsuperscript{1}CREOL, University of Central Florida, Orlando, Florida, USA 32816}
\affiliation{\textsuperscript{2}Department of Applied Physics, Yale University, New Haven, Connecticut, USA 06520}
\begin{abstract}
Both spontaneous emission and resonant energy transfer can be enhanced significantly when the emitter is placed in the vicinity of metallic or crystal structures. This enhancement can be described using the electromagnetic Green tensor and is determined by the dominant surface modes of the structure. 
Here we use the eigenpermittivity formalism to derive the spontaneous emission and FRET rates in the quasistatic regime in a two-constituent
medium with an anisotropic inclusion. We then apply our results to a helical structure supporting synchronous vibrations and evaluate the contribution of these modes, which are associated with a strong and delocalized response. We show that this contribution can result in large Purcell factors and long-range FRET, which oscillates with the helix pitch. These findings may have implications in understanding and controlling the interactions of molecules close to helical structures such as the microtubules.
\end{abstract}

\maketitle

Helical structures like alpha helices, DNA, and microtubules have profound importance in biology. The microtubules (MTs) are composed of identical tubulin-dimer units and therefore they have a regular helical shape, similarly to carbon nanotubes \mbox{\cite{dresselhaus2010perspectives}}. MTs selfassemble from their constituent tubulin-protein units and are critical for the development and maintenance of the cell shape, transport of vesicles, and other components throughout cells, cell signaling, and mitosis. Tubulins have a large dipole moment \mbox{\cite{mershin2004tubulin,preto2015possible,tuszynski2004results,guzman2019tubulin}}  and it was suggested that MT vibrations could generate an electric field in its vicinity \mbox{\cite{tuszynski2016overview,cifra2010electric,thackston2019simulation}}, also beyond the typical Coulomb and van der Waals range.

An optical system can generate a strong electromagnetic field for certain sets of the physical parameters, which are the resonances of the system. 
An eigenvalue is a parameter of the system that corresponds to a resonance, and it can
be obtained by fixing all the other parameters and imposing outgoing boundary conditions without a source. In electrodynamics the eigenvalue is usually defined as an eigenfrequency
or eigenpermittivity of one of the constituents \cite{bergman1980theory,sauvan2013theory,farhi2016electromagnetic}.
In the eigenfrequency formalism a resonance can be approached when
the real physical frequency is close to the usually-complex eigenfrequency. In the eigenpermittivity
formalism a resonance can be achieved when the physical permittivity of one of the constituents is equal to a generally-complex eigenpermittivity of that consistuent.

For a simple system with two uniform and isotropic constituents as in Fig. 1 (a), when the eigenpermittivity ratio $\epsilon_{1m}/\epsilon_2$ is equal to the physical permittivity ratio $\epsilon_{1}/\epsilon_2,$ there is a strong response of the system. In the quasistatic (QS)
regime, in which the typical length scale is much smaller than the
wavelength,  $\epsilon_{1m}/\epsilon_2$ are real and usually negative, see Appendix 1 and Refs. \cite{bergman1979dielectric,bergman1985}. Hence, resonances can usually be approached when the permittivity of one of the constituents is positive and the permittivity of the other is negative, both with low loss. Examples include silver-PMMA \cite{fang2005sub,farhi2016electromagnetic},
silver-water in the high-visible \cite{farhi2017eigenstate}, graphene
\cite{jablan2009plasmonics}, and SiC \cite{hillenbrand2002phonon}.
In the full-Maxwell equation analysis, $\epsilon_{1m}/\epsilon_2$ is usually complex and approaching a resonance requires gain in one of the constituents \cite{bergman1980theory,farhi2016electromagnetic}.

In electrodynamics an eigenstate is usually an electric fields that exists without a source and corresponds to an eigenvalue. Such eigenstates have been used to approximate the field at a resonance in the eigenfrequency formalism \cite{sauvan2013theory} and expand the scattered field in the eigenpermittivity formalism in response to an applied field in a two-constituent medium \cite{bergman1980theory,sauvan2013theory,farhi2016electromagnetic}. These field approximation and field expansion have been generalized to a dipole source excitation independently \cite{sauvan2013theory,farhi2016electromagnetic}, which is of paramount importance for a variety of applications. Another approach for such a calculation is to expand the electric potential of a source in free space according to the inclusion geometry and impose boundary conditions for these modes and the scattered electric potential modes \cite{klimov2004spontaneous}.

Recently, we have shown that in the QS regime when one of the constituents in a two-constituent medium is anisotropic as in Fig. 1 (b), there is an infinite degeneracy of real eigenpermittivities, similarly to the situation in electrodynamics. In this case, however, the eigenperimittivities are real, which can lead to a strong response when an external field is applied. We then used the corresponding eigenfunctions to expand the field in such a setup \cite{farhi2020coupling}.

When the structure is a crystal with a period $a$ as in Fig. 1 (c), one can use an effective $\epsilon\left(\boldsymbol{k}\right)$ when $\lambda\gg a$ \cite{agranovich2013crystal}.
Assuming that $k\approx0$ and $\epsilon\left(\omega\right),$ when the physical frequency is close to a resonant frequency 
$\omega\approx\omega_{T},$ the physical permittivity diverges to plus and minus infinity \cite{kittel1996introduction,hillenbrand2002phonon,carminati2015electromagnetic,joulain2003definition}.
Thus, it can be equal to an eigenpermittivity 
 and result in a strong response. This approximation can also be used in the QS regime when the source-structure distance $l,$ which is on the order of the effective wavelength  \cite{joulain2003definition}, satisfies $ l\gg a.$

The response of a helical structure of Fig. 1 (d) to an incoming electric field
due to the vibrational modes was recently studied in the QS regime.
The arrangement of the units in a \emph{helical periodicity} enabled us to write an effective permittivity  $\epsilon_{1}\left(\boldsymbol{k}\right).$
Then, in order to model axial vibrations we considered an effective permittivity in
the axial axis $\epsilon_{1z}\left(\boldsymbol{k}\right)$ and permittivity value in the other axes $\epsilon_{2},$ the 
permittivity of the host medium. In this work we also investigated the
permittivity when $k\geq2\pi/a,$ where $a$ is the helix pitch. We identified synchronous-vibration modes satisfying
$k=mk_{z}$, where $m,k$ are the cylindrical-mode indices, and $k_{z}=2\pi/a.$
These modes were shown to have $\omega\left(k\right)$ that is close
to real, which is associated with a strong response and delocalization. When the physical frequency $\omega\approx\omega\left(k\right),$
the permittivity is expected to span over a large range of values
and give rise to resonances and delocalization \cite{farhi2020coupling},
similarly to the scenario in crystals mentioned above. Interestingly, delocalized phonons were recently observed in DNAs under physiological conditions \mbox{\cite{gonzalez2016observation}}. 

The local density of states (LDOS) of the electromagnetic field is an important quantity since it determines the magnitude of light-matter interactions such as the spontaneous emission rate. The LDOS is proportional to the imaginary part of the Green tensor, which depends linearly on the electric field generated in response to a dipole excitation \cite{caze2013spatial}. Hence, close to a resonance there is an increase of the scattered field and therefore in the LDOS, which in turn enhances light-matter interactions and spontaneous-emission rate \cite{carminati2006radiative,klimov2004spontaneous,rivera2016shrinking}. To quantify this enhancement one can use the Purcell factor \cite{purcell}, which is defined as spontaneous emission rate in a given system relative to free space.

\begin{figure}[t!]
\begin{centering}
\includegraphics[scale=0.5]{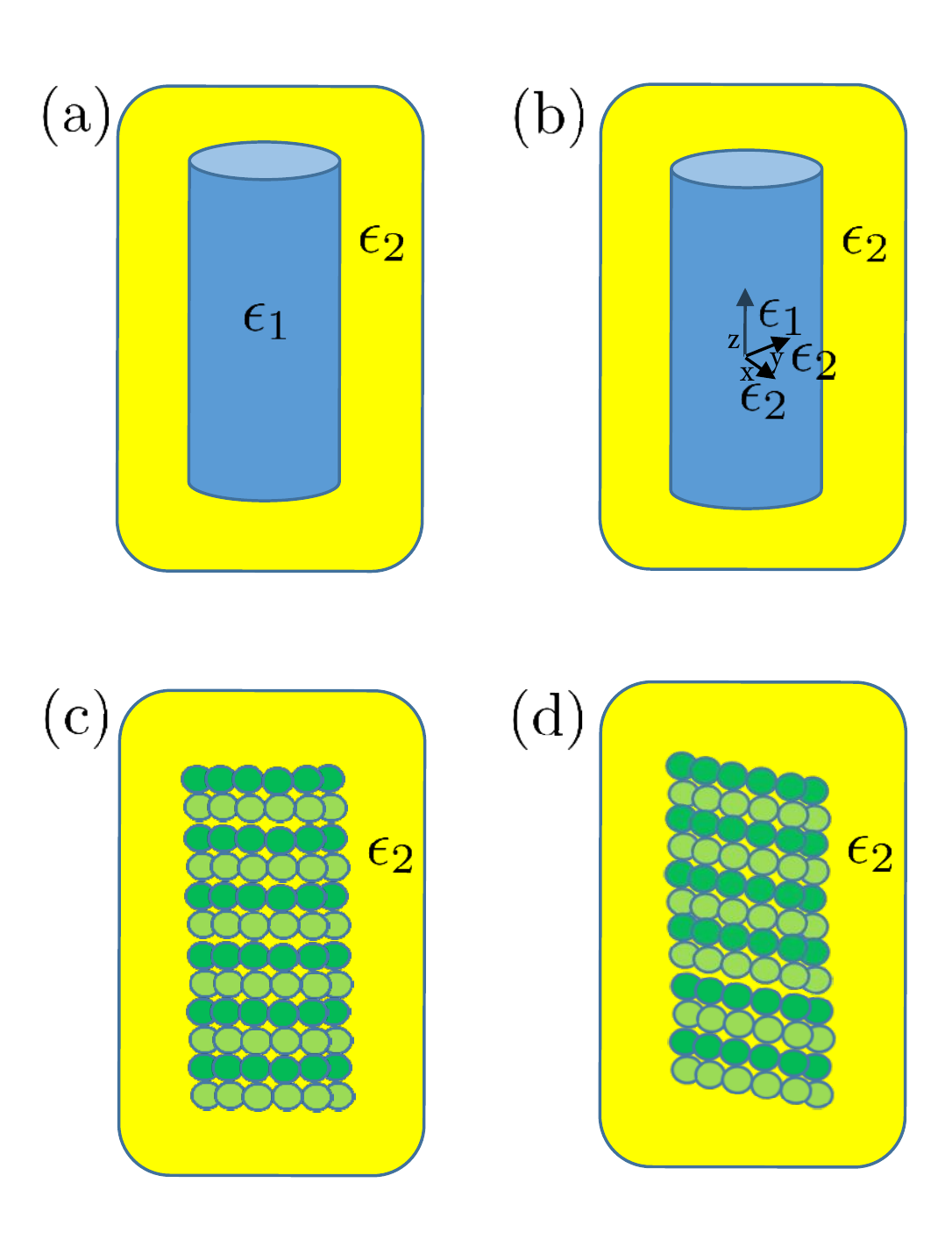}
\par\end{centering}
\caption{(a) Dielectric cylindrical $\epsilon_{1}$ structure in an $\epsilon_{2}$
host medium. (b) Dielectric cylindrical structure with $\epsilon_{1}$
in the $z$ direction and $\epsilon_{2}$ in the $\phi$ and $\rho$
directions, in an $\epsilon_{2}$ host medium. Note that even though the permittivity of the inclusion in the $\phi$ and $\rho$
directions and the permittivity of the host are equal, the different axial permittivity of the inclusion defines an interface. This allows us to model axial vibrations as will be explained.  Periodic longitudinal
(c) and helical (d) arrangements of the constituent units. (c) and
(d) are realizations of (b) with $\epsilon_1 \left(k\right)$ and their vibrational modes are longitudinal
and helical, respectively \cite{kittel1996introduction,farhi2020coupling}. }
\end{figure}

Moreover, when two dipoles are located in proximity to a structure, the Forster Resonance Energy Transfer (FRET) between them is also described in terms of the Green tensor \cite{dung2002intermolecular}. Thus, close to a resonance, we should expect an enhancement in the FRET between the dipoles as well. In free space, such a FRET process between two dipoles is very short range, on the order of 3-4 nanometers. Thus, if a resonant helical structure can mediate FRET between dipoles spaced significantly further apart, it would be of utmost importance in understanding and controlling molecular interactions in the vicinity of such a structure.  

Here we will first evaluate the LDOS and the FRET rate between
two dipoles in the vicinity of an \emph{anisotropic} structure using the eigenpermittivity formalism. We will then apply the results to the case of a generic helical structure supporting axial vibrations and discuss the consequences for strong light-matter interaction, high frequency-selectivity, and structure-mediated long-range energy transfer between dipoles. To the best of our knowledge, this is the first calculation of the interaction between a crystal and a dipole source and between a crystal and electric field with an effective wavelength on the order of the length period of the crystal, which in our case is due to the proximity of the dipole to the structure \mbox{\cite{pendry2000negative}}. For concreteness, we will consider the microtubule, which also has axial periodicity \cite{farhi2020coupling}. 

We start by expanding the electric field using the eigenpermittivity
formalism in the QS regime for a two-constituent system comprising an \emph{isotropic} cylindrical inclusion in a host medium and a point-dipole source situated in the host
medium. An electric potential expansion and Purcell enhancement for such a setup using mode matching were calculated in Ref. \cite{klimov2004spontaneous}. The field generated by the dipole at a position $\boldsymbol{r}$ is
\begin{equation}
E_{\mu}=\sum_{m}\frac{L_{z}}{2\pi}\int dk'\frac{s_{mk'}^{2}}{s-s_{mk'}}\frac{\nabla\psi_{mk'}^{*}\left(\boldsymbol{r}_{0}\right)\cdot\boldsymbol{p}}{\left\langle \psi_{mk'}|\psi_{mk'}\right\rangle }\left(\nabla\psi_{mk'}\left(\boldsymbol{r}\right)\right)_{\mu},
\end{equation}
where $s=\frac{\epsilon_{2}}{\epsilon_{2}-\epsilon_{1}},s_{m}=\frac{\epsilon_{2}}{\epsilon_{2}-\epsilon_{1m}},$
$\boldsymbol{p}$ is the dipole moment, $\boldsymbol{r}_{0}$ is the
dipole location, $\psi_{mk'}$ are the quasielectric potential eigenfunctions, $\mu$ is the
field direction, $\left\langle \psi_{mk'}|\psi_{mk'}\right\rangle =\int\theta_{1}\nabla\psi_{mk'}\cdot\nabla\psi_{mk'}d\boldsymbol{r},$ $\theta_1$ is a window function that equals 1 inside the inclusion volume, and $L_z$ is an arbitrary length that cancels out with $L_z$ in $\left\langle \psi_{mk'}|\psi_{mk'}\right\rangle$ \mbox{\cite{bergman1979dielectric,farhi2016electromagnetic,farhi2017eigenstate,farhi2020coupling}}. Note that when $\epsilon_1\approx \epsilon_{1m}$  there is a large contribution of the corresponding mode in the field expansion.

This formulation was recently generalized to the case of an anisotropic inclusion
by assigning $\epsilon_{1}$ to one axis and $\epsilon_{2}$
to the other two axes to model axial vibrations and we can proceed accordingly with $\epsilon_1\rightarrow\epsilon_{1z}$ and the corresponding eigenfunctions, see Fig. 1 (b) and Ref. \cite{farhi2020coupling}. The Green tensor is proportional to the electric field generated by a dipole and can be expressed as \cite{caze2013spatial} 
\[
G_{\mu\mu'}=\frac{E_{\mu}e_{p\mu'}}{\omega^{2}p}=\frac{E_{\mu}e_{p\mu'}}{k^{2}c^{2}p},\,\,\,k=\frac{2\pi}{\lambda},
\] 
and therefore we readily obtain the the expression for the Green function
\begin{align}
&G_{\mu\mu'}\left(\boldsymbol{r},\boldsymbol{r}_{0}\right)\nonumber \nonumber\\
&=\frac{1}{\omega^{2}}\sum_{m}\frac{L_{z}}{2\pi}\int dk'\frac{s_{mk'}^{2}}{s-s_{mk'}}\frac{\nabla\psi_{mk'}^{*}\left(\boldsymbol{r}_{0}\right)\cdot\boldsymbol{e}_{\mu'}}{\left\langle \psi_{mk'}|\psi_{mk'}\right\rangle }\left(\nabla\psi_{mk'}\left(\boldsymbol{r}\right)\right)_{\mu}.
\end{align}
This expression can then be used to derive to the cross density of states (CDOS) 
\cite{caze2013spatial} \begin{equation}
\overline{\rho}\left(\boldsymbol{r},\boldsymbol{r}_{0}\right)=-\frac{2\omega}{\pi}\mathrm{Im}\left(G_{\mu\mu}\left(\boldsymbol{r},\boldsymbol{r}_{0}\right)\right).
\end{equation}
Similarly, the FRET rate between dipoles at $\boldsymbol{r}_A$ and $\boldsymbol{r}_B$ can be written as \cite{dung2002intermolecular}
\begin{equation}
w_{ab'}^{a'b}=\frac{2\pi}{\hbar^{2}}\left(\frac{\omega_{a'a}}{\epsilon_{0}c^{2}}\right)^{2}\left|\mathbf{p}_{b'b}\overleftrightarrow{G}\left(\boldsymbol{r}_{B},\boldsymbol{r}_{A},\omega_{a'a}\right)\mathbf{p}_{aa'}\right|^2,
\end{equation}
and we arrive at
\begin{align}
&w_{ab'}^{a'b}=\frac{2\pi}{\hbar^{2}}\left(\frac{1}{\epsilon_{0}c^{2}}\right)^{2}\left(\frac{1}{\omega}\frac{L_{z}}{2\pi}\right)^{2}\times\nonumber\\
&\left|\sum_{m}\int dk'\frac{s_{mk'}^{2}}{s-s_{mk'}}\frac{\nabla\psi_{mk'}^{*}\left(\boldsymbol{r}_{B}\right)\cdot\boldsymbol{p}_{aa'}\nabla\psi_{mk'}\left(\boldsymbol{r}_{A}\right)\cdot\boldsymbol{p}_{bb'}}{\left\langle \psi_{mk'}|\psi_{mk'}\right\rangle }\right|^{2}
\end{align}
Assuming a sharp resonance and using the identity $\delta\left(x\right)=\lim_{\epsilon\rightarrow0}\frac{1}{\pi}\frac{\epsilon}{x^{2}+\epsilon^{2}},$
similarly to the spontaneous-emission rate calculation in Ref. \cite{rivera2016shrinking},
one can readily solve analytically the integrals in Eqs. (3) and (5). 

The local density of states (LDOS), which is a private case of the CDOS for $\boldsymbol{r}=\boldsymbol{r}_0$, defined as
\cite{wijnands1997green,economou1983green,joulain2003definition,carminati2015electromagnetic}
\begin{equation}
\overline{\rho}_{\mu}\left(\boldsymbol{r}\right)=-\frac{2\omega}{\pi}\mathrm{Im}\left[G_{\mu\mu}\left(\boldsymbol{r},\boldsymbol{r}\right)\right],
\end{equation}
can also be obtained
\begin{align}
&\overline{\rho}_{\mu}=-\frac{2}{\pi}\frac{1}{\omega}\times\nonumber\\
&\sum_{m}\int dk'\frac{s_{m}^{2}\mathrm{Im}\left(s^{*}\right)}{\left(\mathrm{Re}\left(s\right)-s_{m}\right)^{2}+\left(\mathrm{Im}\left(s\right)\right)^{2}}\frac{\left|\nabla\psi_{m\mu}\right|^{2}}{\tilde{\left\langle \psi_{m}|\psi_{m}\right\rangle }}.
\end{align}
By expressing the spontaneous emission rate in a general setup
\cite{carminati2015electromagnetic}
\begin{equation}
\Gamma=\frac{\pi\omega_{eg}p^{2}}{\hbar\epsilon_{0}}\overline{\rho}_{\mu}\left(\boldsymbol{r},\omega\right)\propto\omega_{eg}\overline{\rho}_\mu,
\end{equation}
where $\omega_{eg}$ is the Bohr frequency between the ground and excited states, 
and the corresponding one in vacuum  \cite{carminati2015electromagnetic}
\begin{equation}
\Gamma_{\mathrm{vacuum}}=\frac{\omega_{eg}^{3}}{3\pi\hbar\epsilon_{0}c^{3}}p^{2},
\end{equation}
we get the following expression for the Purcell factor  
\begin{widetext}
\begin{equation}
\frac{\Gamma}{\Gamma_{\mathrm{vacuum}}}=\frac{3\pi^{2}c^{3}\overline{\rho}_{e,u}\left(\boldsymbol{r},\omega\right)}{\omega_{eg}^{2}}=-3\pi\frac{1}{k^{3}}2\sum_{m}\int dk'\frac{s_{m}^{2}\mathrm{Im}\left(s^{*}\right)}{\left(\mathrm{Re}\left(s\left(k'\right)\right)-s_{m}\left(k'\right)\right)^{2}+\left(\mathrm{Im}\left(s\right)\right)^{2}}\frac{\left|\nabla\psi_{m,k'\mu}\left(\boldsymbol{r}_{0}\right)\right|^{2}}{\tilde{\left\langle \psi_{m}|\psi_{m}\right\rangle }},
\end{equation}
\end{widetext}
We will now examine the specific case of a helical
structure supporting synchronous-vibration modes, which can give rise to resonances. The scattering QS eigenfunctions that correspond to these vibrations satisfy the relation $k=mk_{z}$ due to their functional dependency that is according to the helical symmetry, as illustrated in
Fig. 2 (a) and can be expressed as
\cite{farhi2020coupling} 
\begin{widetext}
\begin{equation}
\psi_{m}=e^{im\left(k_{z}z-\phi\right)}
\left\{ \begin{array}{cc}
A_{4m}K_{m}\left(mk_{z}\rho\right) & \rho>\rho_{2}\\
A_{2m}K_{m}\left(mk_{z}\sqrt{\frac{\epsilon_{1zm}}{\epsilon_{2}}}\rho\right)+A_{3m}I_{m}\left(mk_{z}\sqrt{\frac{\epsilon_{1zm}}{\epsilon_{2}}}\rho\right) & \rho_{1}<\rho<\rho_{2}\\
A_{1m}I_{m}\left(mk_{z}\rho\right) & \rho<\rho_{1}
\end{array}\right.,
\end{equation}
\end{widetext}
where $\rho_{1},\rho_{2}$ are the internal and external inclusion
radii, respectively, and $I_{m},K_{m}$ are the modified Bessel functions. The convergence of Eq. (10) is ensured since $K_{m\geq1}\left(mk\rho\gg a\right)\rightarrow\frac{1}{\sqrt{2mk}}\sqrt{\frac{\pi}{\rho}}e^{-mk\rho},$  $\rho_0>\rho_2,$ and there is always an imaginary part of the permittivity, see Appendix A.1.3.

In crystals, the permittivity is usually expanded in a Fourier series and it couples each field mode with the modes with $\mathbf{k}+\mathbf{G}_n,$ where $\mathbf{G}_n$ is a reciprocal-lattice vector, and there is an effective $\overleftrightarrow{\epsilon_1}(\omega,\mathbf{k})$ that describes the $\omega,\mathbf{k}$ response to an excitation at $\omega,\mathbf{k}$ \mbox{\cite{agranovich2013crystal,yariv1984optical}}.
In our case, the symmetry to discrete translations defines the $k=mk_z$ and $k=nk_z$ modes that represent the ``DC'' and higher-order Fourier components, respectively, see also the static case of electric charges in a helical arrangement in Ref. \cite{18}. Thus, the coupling is to modes with integer multiples of $(\Delta m,\Delta k)=(1,k_z)$ and $\Delta k=nk_z$ apart. At dipole distances on the order of the length period $a,$ the field that is generated by the high-order modes is negligible at the dipole location and therefore the most dominant mode is the $m=1$.

We now analyze classically the vibrational modes that can be excited by the incoming field and generate field as was done in Ref. \mbox{\cite{farhi2020coupling}}.
We consider the coupling of vibrations also to field components with $kc\gg\omega$ that are almost static \mbox{\cite{agranovich2013crystal}} and satisfy $ka\geq1.$ 
When vibrational modes and electric field are coupled they have the same $\omega,\mathbf{k},$ and at low and high $k\mathrm{s},$ $\omega(k)$ of one of the polaritons and the uncoupled vibrational mode are similar \mbox{\cite{kittel1996introduction}}.
We study a structure comprising two types of units with masses $m_1,m_2$ connected by springs $k_1,k_2,k_3,k_4$ as shown in Fig. 2 (b). Denoting the axial displacements of $m_{1,2}$ and the indices of the axial and lateral shifts by $u_{1,2}$ and $s,q$, respectively, and assuming $u_{1,2}=a_{1,2} e^{ikz+im\phi},$ we write the equations of motion (EOM)
\begin{widetext}
\begin{align}
-\omega^{2}m_{1}u_{1sq}=&k_{1}\left(u_{2sq}-u_{1sq}\right)-k_{2}\left(u_{1sq}-u_{2s-1q}\right)-k_{3}\left(u_{1sq}-u_{1sq+1}\right)-k_{3}\left(u_{1sq}-u_{1sq-1}\right),\nonumber\\
-\omega^{2}m_{2}u_{2sq}=&k_{2}\left(u_{1s+1q}-u_{2sq}\right)-k_{1}\left(u_{2sq}-u_{1sq}\right)-k_{4}\left(u_{2sq}-u_{2sq+1}\right)-k_{4}\left(u_{2sq}-u_{2sq-1}\right),
\end{align}
\begin{equation}
\left(\begin{array}{cc}
-\omega^{2}m_{1}+k_{1}+k_{2}+4k_{3}\sin^{2}\left(\left(ka/n-2\pi m/n\right)/2\right) & -\left(k_{2}e^{-ika}+k_{1}\right)\\
-\left(k_{2}e^{ika}+k_{1}\right) & -\omega^{2}m_{1}+k_{1}+k_{2}+4k_{4}\sin^{2}\left(\left(ka/n-2\pi m/n\right)/2\right)
\end{array}\right)\left(\begin{array}{c}
u_{1}\\
u_{2}
\end{array}\right)=\left(\begin{array}{c}
0\\
0
\end{array}\right).
\end{equation}
\end{widetext}

This \emph{1D description} enables us to analyze the behavior of the system in the axial axis while accounting for the lateral interactions in the terms with $k_3,k_4.$ These diagonal terms restrain the movements of $m_1,m_2$ to their sites as in a local oscillator and vanish for the helical functions satisfying $k=mk_z$ (see Eq. (11)). Also, for these modes it can be seen that laterally adjacent units oscillate in-phase. Eq. (13) can be written as $A\mathbf{u}=\omega^2 \mathbf{u},$ where A is a Hermitian matrix and therefore diagonalizable and since $\omega^2$ is real and positive the modes are delocalized. When anharmonicity or dissipation are incorporated, the matrix formulation and Hermiticity no longer hold and localization can arise. We assume that the largest anharmonicity is in the axial forces between lateral units due to the alignment shift of the units upon movement and the distribution of charge along them (see Ref. \mbox{\cite{farhi2020coupling}}, Fig. A1). The anharmonicity in these terms $\propto k_5 u_{1sq}^{2}\left[1-2\cos\left(ka/n-2\pi m/n\right)+\cos\left(2(ka/n-2\pi m/n)\right)\right]$ and translates to an \emph{on-site} anharmonic term, which vanishes for the $k=mk_z$ modes. Moving away from $k=mk_z$ increases the ratio of anharmonicity to dispersion, leading to a more localized response, similarly to interacting diatomic molecules with \emph{internal} anharmonicity \mbox{\cite{kimball1981anharmonicity,hess2000direct}}. From Eq. (13) we calculate $\omega(k)$ for the acoustic and optical modes without anharmonicity. The $k=mk_z$ modes have the same $\omega(k)$ of a 1D crystal (see Fig. 2 (c)) in agreement with the previous analysis in Eq. (11). We then incorporated dissipation into the calculation of  $\omega(k),$ which showed that $\mathrm{Re}(\omega(k))$ 
 is hardly affected and $\mathrm{Im}(\omega(k))$  is constant at all ks, except at large $\gamma\mathrm{s}$ that suppress the acoustic modes, see Ref. \mbox{\cite{farhi2020coupling}}, Appendix B3.
\begin{figure}
\begin{centering}
\includegraphics[scale=0.5]{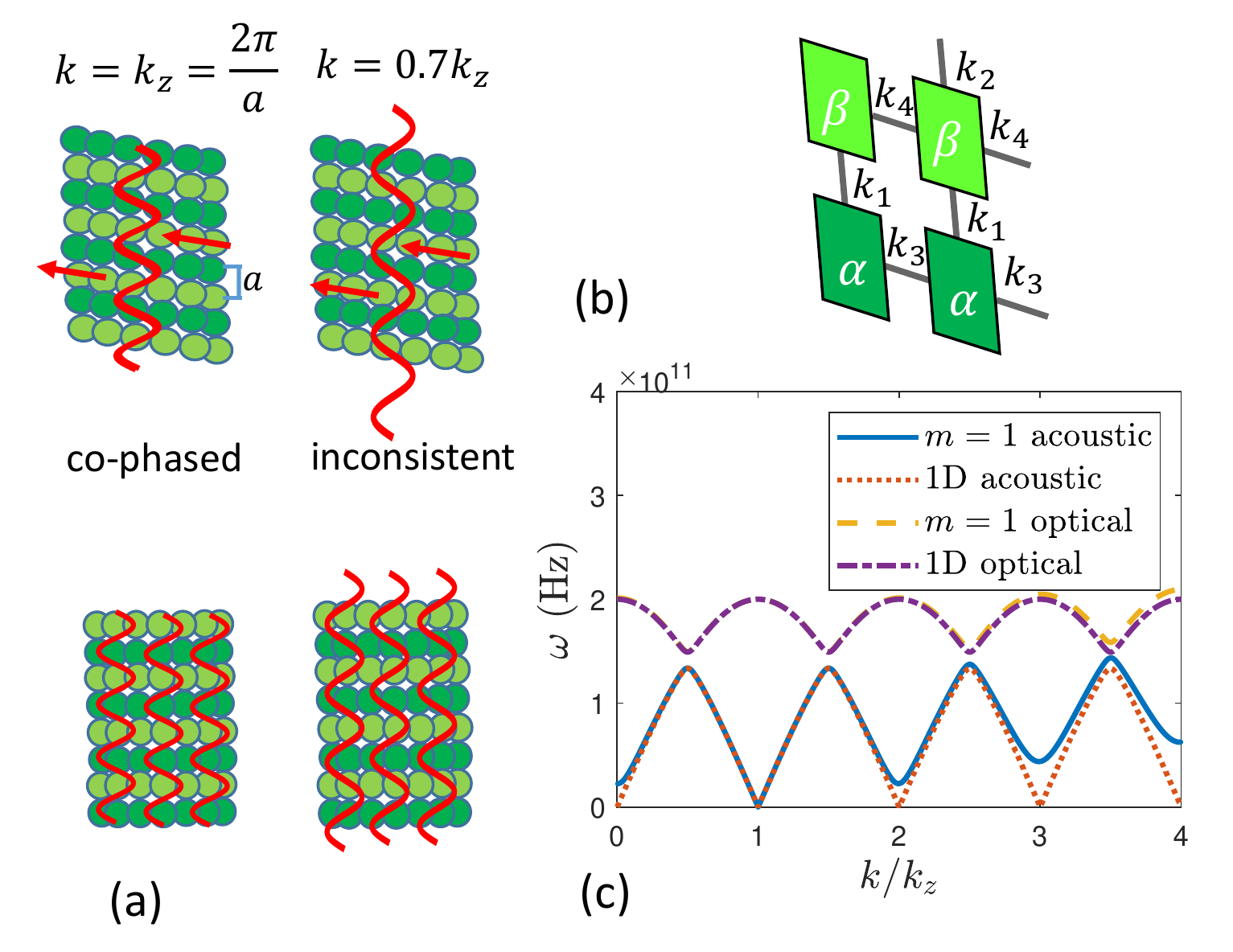}
\par\end{centering}
\caption{Vibrational-mode analysis for a helical structure.(a) The illustrations
show that $k=k_{z}m$ are allowed when requiring decoupling between
the axial protofilaments. (b) The structure is composed of two units
denoted by $\alpha,\beta$ with masses $m_{1},m_{2}$ connected with
springs $k_{1},k_{2},k_{3},k_{4}$. (c) $\omega(k)$ for the acoustic
and optical $m=1$ helix and 1D crystal modes. The microtubule parameters
are $m_{1}=m_{2}=0.9\cdot10^{-22}\,(\mathrm{Kg}),k_{1}=0.8,k_{2}=0.1,k_{3}=k_{4}=0.2\,(\mathrm{N/m}),$
where $k_{4}$ is of the order of magnitude of the value in Ref. \cite{portet2005elastic}.}
\end{figure}
The physical permittivity can then be written similarly to the derivation for a harmonic oscillator \mbox{\cite{kittel1996introduction}} where the oscillator eigenfrequency $\omega_T \rightarrow \omega (\mathbf{k})$ \cite{farhi2020coupling}
\begin{equation}
\epsilon_{1}=1+\frac{4\pi Nq^{2}}{m_{r}\left[\omega^{2}\left(\boldsymbol{k}\right)-\omega^{2}\right]},
\end{equation}
where $q$ is the unit charge, $m_{r}$ is the effective mass, and
$N$ is the charge density. Note that in electrodynamics in the quasistatic regime, the dependency on $\omega$ is negligible and therefore the polariton eigenfrequency is approximately determined by the vibrational modes.

We are now in position to derive the LDOS and the FRET rate for helical structures using the expressions in Eqs. (5) and (7). For simplicity we focus on the $m=1$ modes, which dominate at large distances, and proceed without anharmonic
terms \cite{farhi2020coupling}, similarly to Ref. \cite{kittel1996introduction}. We first calculate $\epsilon_{1}\left(k\right)$
and $\epsilon_{1k}$ using the expression above and the boundary conditions
\cite{farhi2020coupling}, respectively, to observe the intersection
points between them, which are the resonances. In the calculation of  $\epsilon_{1}\left(k\right)$ we chose $q=12e,$
where $e$ is the electron charge \cite{van2007electrophoresis},
and spring constants on the order of the one reported in Ref. \cite{de2003deformation}. We also
use in the expression of $\epsilon_{1}(k),$   $\omega^{2}\left(\boldsymbol{k}\right)$
that incorporates dissipation and has a constant imaginary part \cite{farhi2020coupling}. To compare the LDOS and FRET results to an isotropic dielectric structure (which is the standard modeling of helical structures of this kind), with $\epsilon_{1}=1.5+0.1i$, we set $\mathrm{Im}\left(\omega_{n}\right)=\frac{\pi Nq^{2}}{4m_{r}},$ which
 satisfies $\mathrm{Im}\left(\epsilon_{1}\right)=0.1i$ 
at an intersection point. Then, using $\epsilon_{1}\left(k\right)$
and $\epsilon_{1k},$  we perform the calculation of the LDOS and FRET for the helical structure setup and compare the results. For details about the calculations for the isotropic structure see Appendix 2.

In Fig. 3 we present the physical permittivity $\epsilon_{1}$ and the eigenpermittivities $\epsilon_{1k}$ for the
first few modes of the helical structure. Due to the anisotropy there
is an infinite number of eigenpermittivities for a given $k$ value,
unlike the case of an isotropic medium. While this resembles electrodynamics, in which there are are multiple resonances at a given $k$ value,
the eigenpermittivities in this case are real and can give rise
to a strong response, especially for the first resonances where $\mathrm{Im}\left(\epsilon_{1}\right)$
is small. Since the resonances are discrete, if we assume that each resonance is a continuous function of $\omega,k,$  when varying $\omega$ we will encounter closely spaces resonances (one can think of resonances represented by e.g., parallel diagonal lines in  $\omega,k$). This is in qualitative agreement with the closely-space resonances in frequency in the experimental results in Ref. \cite{sahu2013atomic}. 

\begin{figure}[h]
\begin{centering}
\includegraphics[width=8cm]{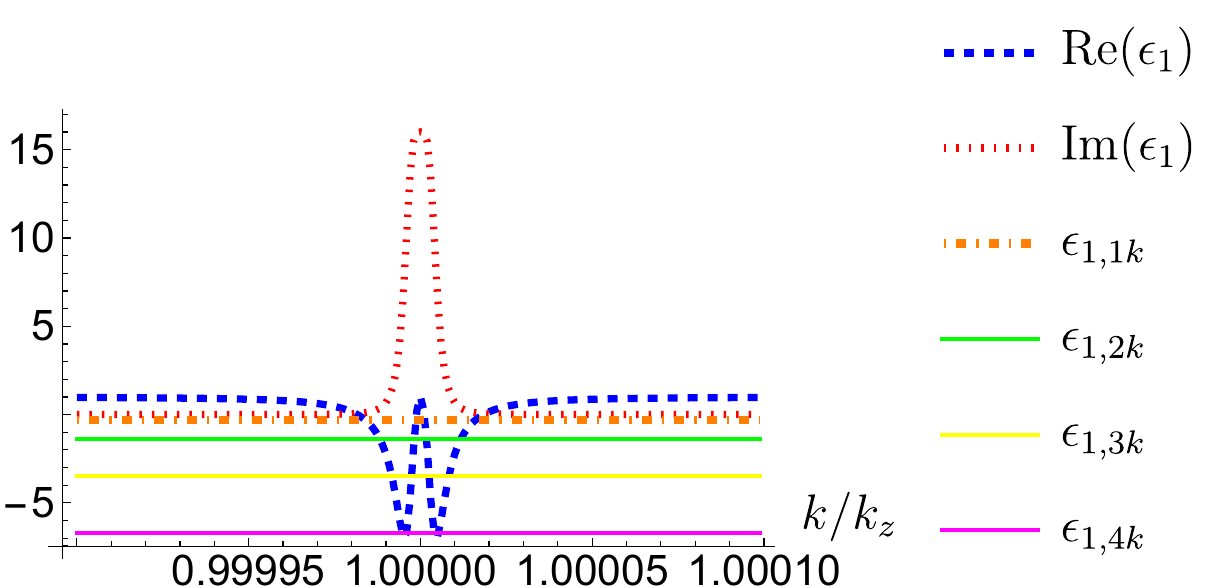}
\par\end{centering}
\caption{Physical permittivity $\epsilon_1(k)$ and first eigenpermittivities $\epsilon_{1,n\, k}$ of the helical
structure, where we used 
the parameters of Fig. 2 and $\rho_{1}=7,\rho_{2}=12\mathrm{nm}$. The real parts of the physical
permittivity and the first eigenpermittivity intersect at two $k$
values with $\mathrm{Im}\left(\epsilon_{1}\right)=0.1i,$ resulting
in large contributions of the corresponding eigenfunctions. Note that we consider a single $\omega,$ and $k$ in the QS expansion can take any value and is not required to satisfy $kc/n=\omega.$}
\end{figure}

In Fig. 4 we present the Purcell factors in the $z$ and $\rho$ directions, which are the dominant ones, as functions of $\rho$ for the helical
and isotropic structures, both with $\rho_{1}=7\mathrm{nm},\rho_{2}=12\mathrm{nm}.$
It can be seen that for the isotropic structure the magnitudes in the $\rho$ direction are larger, whereas for the helical structure the magnitudes in the $\rho$ and $z$ directions are similar. Interestingly, close to the structure the LDOS of the helical structure
dominates since the $m=1$ modes that extend the farthest have a larger response compared to the isotropic structure, 
 while at some distance away the response of the isotropic
structure is larger. This can be explained by the strong response
of the helical structure up to an interaction distance
on the order of $a$ due to the synchronous-vibration modes, which depend on $\rho$ via $K(k_z\rho),$ see Eq. (11). In addition, the modes with larger interaction distances, which have small $k\mathrm{s},$ 
are present in the isotropic structure but not in the helical structure
since $\epsilon_1\approx1 \Rightarrow 1/(s-s_k)\approx 0,$ and therefore they have a negligible contribution in the field expansion, see Eq. (2). This may suggest two distinct mechanisms of interaction in these regions. Finally, we note that Purcell factors depend on the frequency as follows $\Gamma/\Gamma_0\propto 1/\omega^3,$ see Eq. (10). Since in our case  $\omega=2\cdot 10^{11}\,\mathrm{(1/s)},$ it gives e.g., an additional factor of $5\cdot 10^8$ compared to the calculation in the near infrared in Ref. \mbox{\cite{rivera2016shrinking}, Fig. 3}.

In Fig. 5 we present the normalized FRET rates
between two axially-distanced dipoles oriented in the $z$ direction at a distance $a$ from the helical 
and isotropic structures as well as in free space. The isotropic structure exhibits a larger FRET range compared to free space, similarly to a Gaussian beam in which the modes interfere. Importantly, at a given time the FRET close to the helical structure setup has an approximately constant amplitude and oscillates with a period $a,$ due to the $\exp(ik_zz)$ dependency in Eq. (11). Note that since we have incorporated dissipation into the permittivity, the FRET rate in this model decays in space, at a distance that is larger than the one displayed in the graph and can be approximated using $\Delta x\approx1/(\sqrt{2}\Delta k),$ where $\Delta k$ is according to the integrand in Eq. (5).  We also calculated the FRET rates for dipoles oriented along the $\rho$ and $\phi$ directions and, in the helical-structure setup, they have approximately the same $z$ dependency since the contribution is dominated by $\exp(ik_z z),$ see Eq. (11), and their relative magnitudes are 1.14 and $1.64\cdot 10^{-5},$ respectively. The scaling of the FRET as a function of the dipole radius for dipoles oriented along $z,\rho,\phi$ at distances on the order of $a$ or larger can be approximated as $\left(k_{z}K_{1}\left(k_{z}\rho\right)\right)^{2},\left(K_{1}'\left(k_{z}\rho\right)\right)^{4},\left(\frac{1}{\rho}K_{1}\left(k_{z}\rho\right)\right)^{4},$ respectively.  
Finally, the FRET rate as a function of $z$ between dipoles oriented along $z$ and $\rho$ in the helical-structure setup is shifted in phase by $\pi/2,$ which implies that for dipoles at the same $z$ location the FRET rate will be maximal when they are parallel. Incorporating the induced electric response and the anharmonic terms is expected to result in a shorter FRET distance close to the helical structure. Clearly, the effect of including the anharmonic terms depends on their coefficients and the strength of the incoming field, which will determine the mode amplitude. Usually, on a resonance since the mode amplitude is large and the anharmonic terms are significant, it will increase imaginary part of $\omega_k,$ which in turn will reduce the strength and axial extent of the response since $\omega$ is real. However, assuming that the dominant anharmonicity is in the axial forces between lateral units, this effect is expected to be dominant only away from the modes satisfying $k=mk_z,$ where this anharmonicity is large. Thus, excitation at an $\omega$ that is significantly different than $\omega_k(k=mk_z)$ will result in a weaker and more localized response. Additional anharmonic terms can decrease the FRET range. 

\begin{figure}[h]
\begin{centering}
\includegraphics[width=8cm]{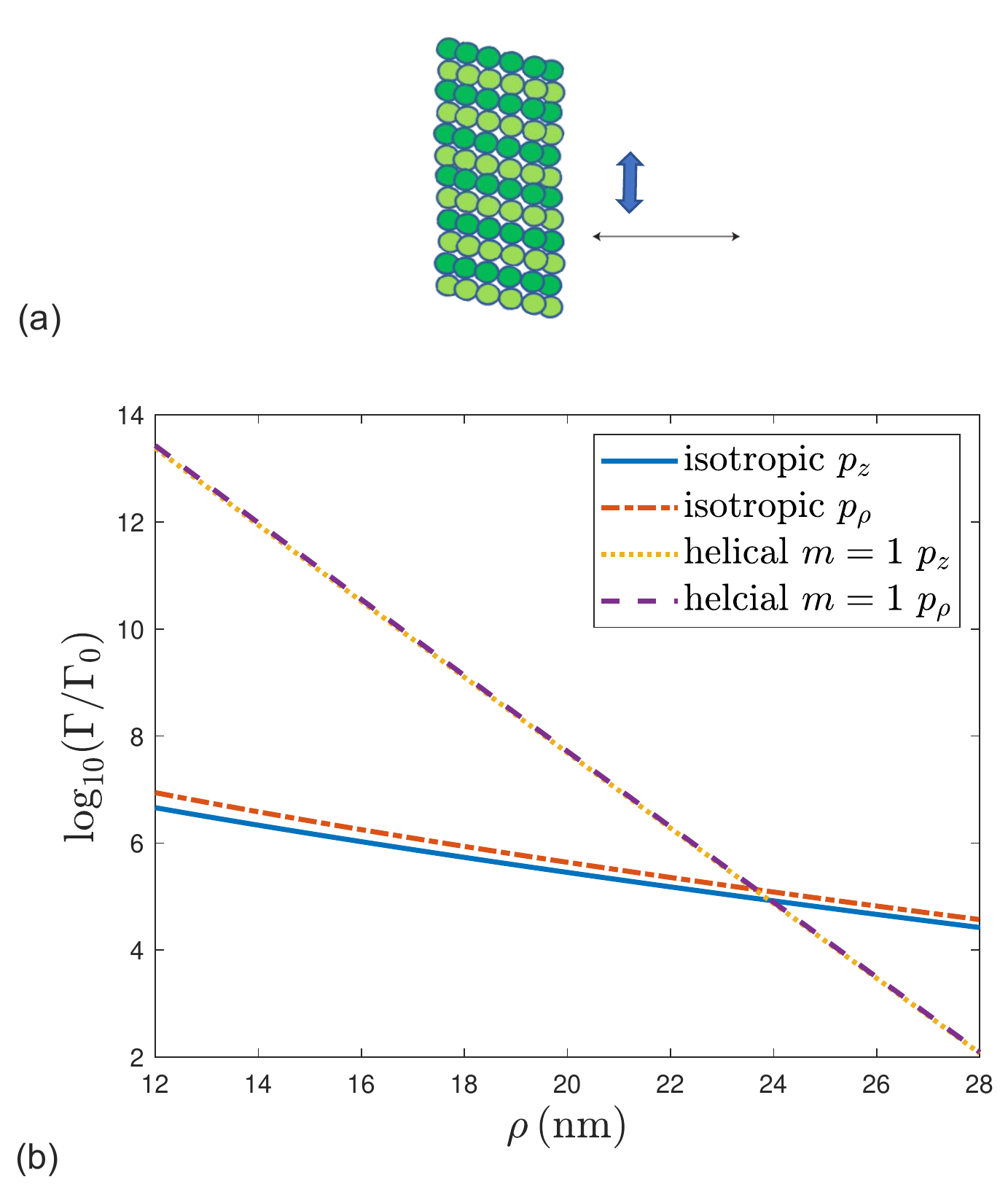}
\par\end{centering}
\caption{ (a) A setup of a dipole in proximity to a helical structure where the spontaneous emission rate of the dipole is enhanced due to the proximity to the structure (b) Purcell factors in the $z$ and $\rho$ directions as functions of the radius for an isotropic and helical
structures with $\rho_{1}=7\mathrm{nm}$
and $\rho_{2}=12\mathrm{nm},$ where the helical structure has $a=8\mathrm{nm}.$ The Purcell factors in the $\phi$ direction are smaller by at least an order of magnitude since $1/\rho<k,$ except at close distances to the isotropic structure, where the magnitudes are negligibible compared with the helical structure. The Purcell factors
of the helical structure are dominant for $12\mathrm{nm}<\rho<24\mathrm{nm}$
and the ones of the isotropic structure are dominant for
$\rho>24\mathrm{nm}.$ In addition, the Purcell factors in the $\rho$ direction are larger for the isotropic structure whereas the magnitudes in the $\rho$ and $z$ directions are similar for the helical structure.}
\end{figure}

\begin{figure}[t]
\begin{centering}
\includegraphics[width=8cm]{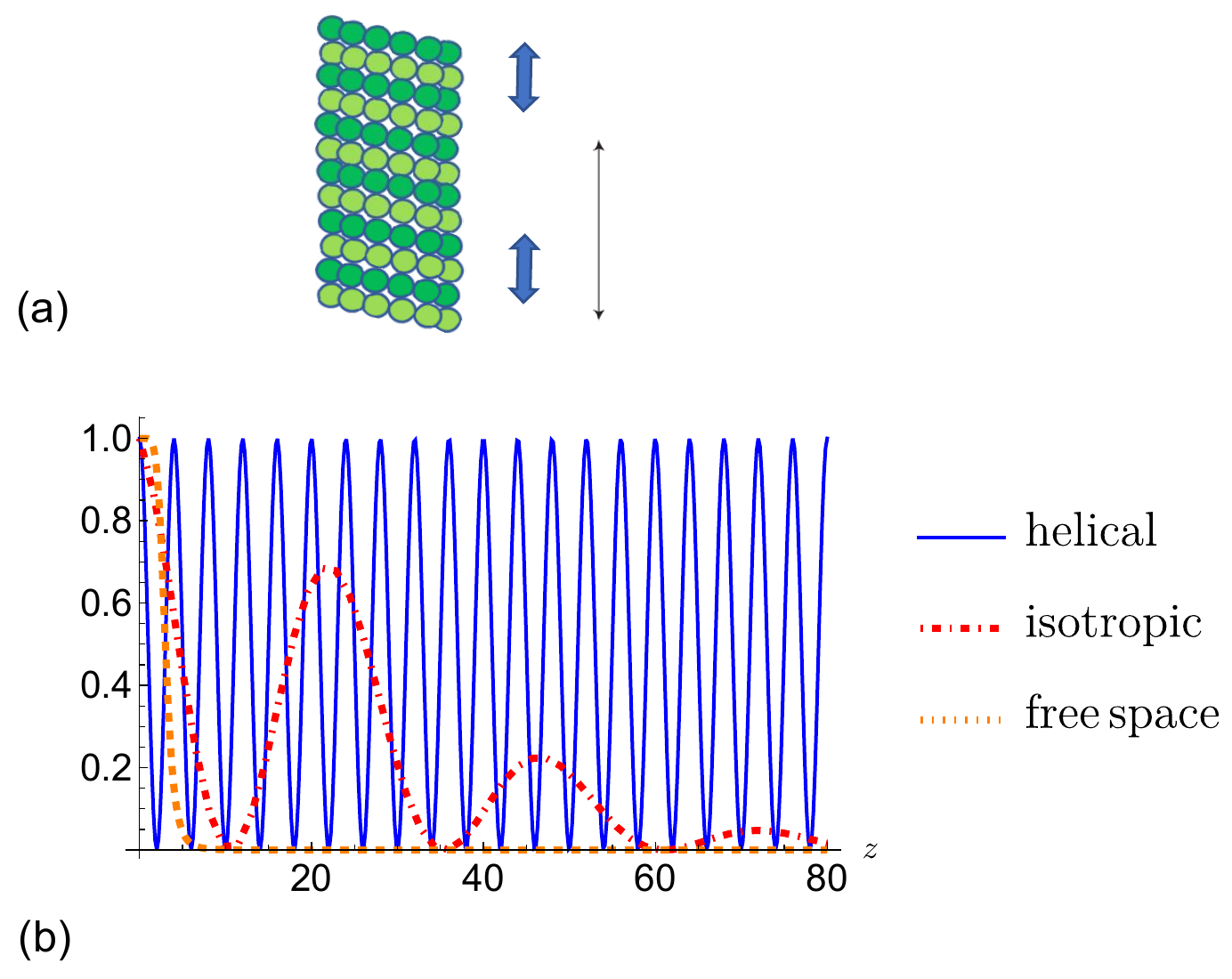}
\par\end{centering}
\caption{(a) The setup of two dipoles in proximity to the helical structure, which can transfer energy via the structure. (b) Normalized FRET rates as functions of the axial distance between two dipoles oriented in the $z$ direction 
for an isotropic dielectric structure with $\epsilon_{1}=1.5+0.1i$,
helical structure, and free space. The distance of the dipoles from the structures is $a,$ which is the helix pitch.}
\end{figure}

%

In conclusion, we first derived the density of states, Purcell factors,
and FRET rate in the eigenpermittivity formalism for a two-constituent
system with isotropic and anisotropic inclusions. We then applied this formulation to the case of a helical structure supporting
axial vibrations and compared it with an isotropic dielectric structure. We showed
that the helical structure can greatly enhance the spontaneous emission
rate up to distances on the order of the helix pitch and that at much larger distances
the dielectric response dominates. Finally, we showed that helical
structures can mediate long-range FRET between two dipoles. This could be crucial for understanding and controlling molecular interactions in the vicinity of such structures. Our results may be of particular relevance for phenomena associated with biological helical structures such DNAs, microtubules, and alpha helices, and could relate to fundamental questions in biology such as the role of electrodynamics in explaining long-range interactions and synchronization between distant molecules.

\section*{Appendix}
\subsection*{A.1.1 Expansion of the potential of a dipole for an anisotropic and spatially-dispersive
inclusion}
We will start by expanding the physical potential of a charge distribution
in a two-constituent medium, in which both constituents are isotropic
and spatially uniform, similarly to the treatment in Refs. \mbox{\cite{bergman1979dielectric,bergman2014,farhi2016electromagnetic}}.
We will then develop an expansion for an inclusion with an anisotropic
and spatially-uniform permittivity and simplify it for a dipole source.
Finally, we will formulate the field expansion for a $k$-dependent
inclusion permittivity where the modes are uncoupled and analyze the
scattered field for a crystal inclusion.

In the quasistatic regime we use Poisson's equation in a two-constituent
medium for the electric potential of a charge distribution $\tilde{\rho}\left(\mathbf{r}\right).$
When both constituents have a spatially uniform and isotropic permittivities
we write \mbox{\cite{bergman1979dielectric,bergman2014,farhi2016electromagnetic}}
\begin{gather}
\nabla\epsilon\nabla\psi=\tilde{\rho}\left(\mathbf{r}\right),\nonumber\\
\nabla^{2}\psi\left(\mathbf{r}\right)=\nabla\cdot\theta_{1}\left(\mathbf{r}\right)u\nabla\psi\left(\mathbf{r}\right)+\tilde{\rho}\left(\mathbf{r}\right)/\epsilon_{2},\,\,\,\,u\equiv\frac{\epsilon_{2}-\epsilon_{1}}{\epsilon_{2}},\nonumber
\end{gather}
where $\theta_{1}\left(\mathbf{r}\right)$ is a window function that
equals 1 inside the inclusion, $\text{\ensuremath{\epsilon}}_{1}$
is the inclusion permittivity, and $\epsilon_{2}$ is the host-medium
permittivity. The potential can be regarded as generated by the external
charge distribution $\tilde{\rho}\left(\mathbf{r}\right)/\epsilon_{2}$
and $\nabla\cdot\theta_{1}\left(\mathbf{r}\right)u\nabla\psi\left(\mathbf{r}\right).$
Therefore, it can also be expressed as $\psi=\psi_{0}+\psi_{\mathrm{sc}}$
in terms of the potential $\psi_{0}$ generated by the charge distribution
in a uniform $\epsilon_{2}$ medium and $\psi_{\mathrm{sc}}$ that is generated due
to the existence of the inclusion.

An eigenstate $\psi_{n},$ which exists in a system without a source,
is defined as follows 
\begin{gather}
\nabla^{2}\psi_{n}\left(\mathbf{r}\right)=\nabla\cdot\theta_{1}\left(\mathbf{r}\right)u_{n}\nabla\psi_{n}\left(\mathbf{r}\right),\ \ \ u_{n}\equiv\frac{\epsilon_{2}-\epsilon_{1n}}{\epsilon_{2}},\nonumber\\
\psi_{n}\left(\mathbf{r}\right)=\int G\left(\mathbf{r}-\mathbf{r}^{'}\right)\nabla\cdot\theta_{1}u_{n}\nabla\psi_{n}\left(\mathbf{r}^{'}\right)d\mathbf{r}'\nonumber\\
=u_{n}\int\theta_{1}\left(\mathbf{r}^{'}\right)\nabla G\left(\mathbf{r}-\mathbf{r}^{'}\right)\nabla\psi_{n}\left(\mathbf{r}'\right)d\mathbf{r}',\nonumber
\end{gather}
where $G\left(\mathbf{r}-\mathbf{r}'\right)$ is Green's function
of Poisson's equation and we performed integration by parts. We define
the operator $\ensuremath{\hat{\Gamma}}$ as 
\[
\hat{\Gamma}\psi_{n}=\int\theta_{1}\left(\mathbf{r}^{'}\right)\nabla G\left(\mathbf{r}-\mathbf{r}^{'}\right)\nabla\psi_{n}\left(\mathbf{r}^{'}\right)d\mathbf{r}'
\]
 and write
\[
\psi_{n}=u_{n}\hat{\Gamma}\psi_{n},\,\ s_{n}\psi_{n}=\hat{\Gamma}\psi_{n},\ \ s_{n}=\frac{1}{u_{n}}.
\]
Since $\hat{\Gamma}$ is self adjoint all the eigenvalues $s_n$ are real. In addition, at the large $n$ limit $s_n=1/2$ and therefore $\epsilon_1/\epsilon_2=-1$ is an accumulation point of the eigenperimittivity ratios \cite{bergman1979dielectric,bergman1985,bergman2014perfect}. 
We then obtain \cite{bergman1979dielectric,bergman2014,farhi2016electromagnetic}
\begin{align}
\psi&=u\hat{\Gamma}\psi+\psi_{0}\nonumber\\
&=\frac{1}{1-u\hat{\Gamma}}\psi_{0}=\psi_{0}+\frac{u\hat{\Gamma}}{1-u\hat{\Gamma}}\psi_{0} \nonumber\\
&=\psi_{0}+\sum_{n}\frac{u\hat{\Gamma}}{1-u\hat{\Gamma}}\left|\psi_{n}\right\rangle \left\langle \psi_{n}|\psi_{0}\right\rangle \nonumber\\
&=\psi_{0}+\sum_{n}\frac{s_{n}}{s-s_{n}}\left|\psi_{n}\right\rangle \left\langle \psi_{n}|\psi_{0}\right\rangle. \nonumber
\end{align}
By using  \cite{farhi2016electromagnetic,farhi2017eigenstate} 
\begin{gather}
\left\langle \psi_{n}|\psi_{0}\right\rangle =\int d\mathbf{r}\theta_{1}\nabla\psi_{n}^{*}\cdot \nabla\psi_{0}\nonumber\\
=\int d\mathbf{r}\theta_{1}\left(\mathbf{r}\right)\nabla\psi_{n}^{*}\left(\mathbf{r}\right)\cdot \nabla\int G\left(\mathbf{r}-\mathbf{r}'\right)q\delta\left(\mathbf{r}'-\mathbf{r}_{0}\right)d\boldsymbol{r}' \nonumber\\
=qs_{n}\int d\mathbf{r}'\psi_{n}^{*}\left(\mathbf{r}'\right)q\delta\left(\mathbf{r}'-\mathbf{r}_{0}\right)=qs_{n}\psi_{n}^{*}\left(\mathbf{r}_{0}\right). \nonumber
\end{gather}
we get for a point charge
$$\psi=\psi_{0}+q\sum_{n}\frac{s_{n}^{2}}{s-s_{n}}\left|\psi_{n}\right\rangle \psi_{n}^{*}\left(\mathbf{r}_{0}\right) \nonumber.$$
The eigenstates are assumed to be normalized, where the inner product
is defined as
\[
\ensuremath{\left\langle \psi_{n}|\psi_{n}\right\rangle =\int d\mathbf{r}\theta_{1}\nabla\psi_{n}^{*}\cdot\nabla\psi_{n}}.
\]
Now we develop the expansion of the potential for an anisotropic inclusion  
permittivity as was done in Ref. \cite{farhi2020coupling}. We denote the inclusion permittivity tensor by $\overleftrightarrow{\epsilon}_{1}$
and write 
\begin{gather}
\nabla\overleftrightarrow{\epsilon}\nabla\psi=\tilde{\rho}\left(\mathbf{r}\right),\nonumber\\
\epsilon_{2}\nabla^{2}\psi+\nabla\theta_{1}\left(\overleftrightarrow{\epsilon}_{1}-\epsilon_{2}\right)\nabla\psi=\tilde{\rho}\left(\mathbf{r}\right),\nonumber\\
\epsilon_{2}\nabla^{2}\psi=\frac{\tilde{\rho}\left(\mathbf{r}\right)}{\epsilon_{2}}\mathbf{+}\nabla\theta_{1}\frac{\left(\epsilon_{2}-\overleftrightarrow{\epsilon}_{1}\right)}{\epsilon_{2}}\nabla\psi,\nonumber\\
\nabla^{2}\psi\left(\mathbf{r}\right)=\nabla\cdot\theta_{1}\left(\mathbf{r}\right)\overleftrightarrow{u}\nabla\psi\left(\mathbf{r}\right)+\frac{\tilde{\rho}\left(\mathbf{r}\right)}{\epsilon_{2}},\,\,\,\overleftrightarrow{u}\equiv\frac{\epsilon_{2}I-\overleftrightarrow{\epsilon}_{1}}{\epsilon_{2}}.\nonumber
\end{gather}
where $I$ is the unit matrix.

We define an eigenfunction $\ensuremath{\psi_{k}}$ as follows
\begin{gather}
\psi_{k}\left(\mathbf{r}\right)=\int G\left(\mathbf{r}-\mathbf{r}^{'}\right)\nabla\cdot\theta_{1}\overleftrightarrow{u}\nabla\psi_{k}\left(\mathbf{r}'\right)d\mathbf{r}'\nonumber\\
=\int G\left(\mathbf{r}-\mathbf{r}^{'}\right)\frac{\partial}{\partial'i}\theta_{1}u_{k,ij}\frac{\partial}{\partial'j}\psi_{k}\left(\mathbf{r}'\right)d\mathbf{r}'\nonumber\\
=\sum_{i,j}u_{ij,k}G\left(\mathbf{r}-\mathbf{r}^{'}\right)\frac{\partial}{\partial'}_{i}\theta_{1}\left(\boldsymbol{r}'\right)\frac{\partial}{\partial'_{j}}\psi_{k}\left(\mathbf{r}'\right)d\mathbf{r}'\nonumber\\
=\sum_{i,j}u_{ij,k}\theta_{1}\left(\boldsymbol{r}'\right)\frac{\partial}{\partial'}_{i}G\left(\mathbf{r}-\mathbf{r}^{'}\right)\frac{\partial}{\partial'_{j}}\psi_{k}\left(\mathbf{r}'\right)d\mathbf{r}',\nonumber
\end{gather}
where we performed integration by parts and $\,u_{ij}\equiv\delta_{ij}-\frac{\epsilon_{1ij}}{\epsilon_{2}}$.

For a diagonal form of $\overleftrightarrow{\epsilon}$ we have
\begin{gather}
\psi_{k}\left(\mathbf{r}\right)=u_{i,k}\int G\left(\mathbf{r}-\mathbf{r}'\right)\frac{\partial}{\partial'_{i}}\theta_{1}\left(\mathbf{r}'\right)\frac{\partial}{\partial'_{i}}\psi_{k}\left(\mathbf{r}'\right)d\mathbf{r}'\nonumber\\
=\sum_{i}u_{i,k}\int\theta_{1}\left(\mathbf{r}'\right)\frac{\partial}{\partial'_{i}}G\left(\mathbf{r}-\mathbf{r}'\right)\frac{\partial}{\partial'_{i}}\psi_{k}\left(\mathbf{r}'\right)d\mathbf{r}'.\nonumber
\end{gather}
For $\left(\epsilon_{x},\epsilon_{y},\epsilon_{z}\right)=\left(\epsilon_{2},\epsilon_{2},\epsilon_{1z}\right)$ we get
\begin{gather}
\psi_{k}\left(\mathbf{r}\right)=u_{zk}\int G\left(\mathbf{r}-\mathbf{r}^{'}\right)\frac{\partial}{\partial'_{z}}\theta_{1}\left(\mathbf{r}^{'}\right)\frac{\partial}{\partial'_{z}}\psi_{k}\left(\mathbf{r}^{'}\right)d\mathbf{r}^{'}\nonumber\\
=u_{zk}\int\theta_{1}\left(\mathbf{r}^{'}\right)\frac{\partial}{\partial'_{z}}G\left(\mathbf{r}-\mathbf{r}^{'}\right)\frac{\partial}{\partial'_{z}}\psi_{k}\left(\mathbf{r}^{'}\right)d\mathbf{r}^{'},\nonumber
\end{gather}
and write the eigenvalue equation
\begin{gather}
\psi_{k}=u_{zk}\hat{\Gamma}_{z}\psi_{k},\,\,\,s_{zk}\psi_{k}=\hat{\Gamma}_{z}\psi_{k},\nonumber\\
s_{zk}=1/u_{zk}=\epsilon_{2}/\left(\epsilon_{2}-\epsilon_{1zk}\right),\nonumber
\end{gather}
where $s_{zk}$ is an eigenvalue. Note that here the physical permittivity
of the inclusion $\epsilon_{1}$ is spatially uniform and the index
$k$ denotes the mode index. Similarly, we write the expansion of
$\psi$ for this case 
\[
\psi=\psi_{0}+\sum_{k}\frac{s_{zk}}{s_{z}-s_{zk}}\left|\psi_{n}\right\rangle \left\langle \psi_{n}|\psi_{0}\right\rangle.
\]
For a point charge we substitute the eigenvalue equation in the inner
product to obtain
\begin{gather}
\left\langle \psi_{k}|\psi_{0}\right\rangle =\int d\mathbf{r}\theta_{1}\left(\mathbf{r}\right)\frac{\partial}{\partial z}\psi_{k}^{*}\left(\mathbf{r}\right)\frac{\partial}{\partial z}\psi_{0}\left(\mathbf{r}\right)\nonumber\\
=\frac{4\pi}{\epsilon_{2}}\int d\mathbf{r}\theta_{1}\left(\mathbf{r}\right)\frac{\partial}{\partial z}\psi_{k}^{*}\left(\mathbf{r}\right)\frac{\partial}{\partial z}G\left(\mathbf{r}-\mathbf{r}^{'}\right)*q\delta\left(\mathbf{r}^{'}-\mathbf{r}_{0}\right)\nonumber\\
=\frac{4\pi q}{\epsilon_{2}}s_{zk}\psi_{k}^{*}\left(\mathbf{r}_{0}\right).\nonumber
\end{gather}
We then consider a dipole composed of two charges and write
\begin{gather}
\left\langle \psi_{k}|\psi_{0}\right\rangle =s_{zk}q\left(\psi_{k}^{*}\left(\mathbf{z}_{0}+\mathbf{d}/2\right)-\psi_{k}^{*}\left(\mathbf{z}_{0}-\mathbf{d}/2\right)\right)\nonumber\\
=s_{zk}qd\frac{\left(\psi_{k}^{*}\left(\mathbf{z}_{0}+\mathbf{d}/2\right)-\psi_{k}^{*}\left(\mathbf{z}_{0}-\mathbf{d}/2\right)\right)}{d}.\nonumber
\end{gather}
For a cylindrical inclusion, the eigenfunctions have two indices $m,k.$
All in all, we obtain for $\psi$ 
\begin{equation}
\psi=\psi_{0}+\frac{4\pi}{\epsilon_{2}}\sum_{m}\int\frac{s_{zk}^{2}}{s_{z}-s_{zk}}\left|\psi_{m,k}\right\rangle \nabla\psi_{m,k}^{*}\left(\mathbf{r}_{0}\right)\cdot\mathbf{p}dk,\,\,\,
\end{equation}
where the inner product for the normalization is
\[
\ensuremath{\left\langle \psi_{k}|\psi_{k}\right\rangle =\int d\mathbf{r}\theta_{1}\left(\mathbf{r}\right)\frac{\partial}{\partial z}\psi_{m,k}^{*}\left(\mathbf{r}\right)\frac{\partial}{\partial z}\psi_{m,k}\left(\mathbf{r}\right)}.
\]
We now formulate an expansion for a $k$-dependent inclusion permittivity
without coupling between modes. This is the situation in an electron
gas, where the physical permittivity value is associated with each
mode \cite{kittel1996introduction}. We first write the response of the inclusion to an
excitation at a given $k$
\[
\psi_{\mathrm{sc},k}=\frac{u_{zk}\Gamma_{z}}{1-u_{zk}\Gamma_{z}}\psi_{0k},
\]
where $u_{zk}$ corresponds to the physical inclusion permittivity
at a given $k$ and 
\[
\ensuremath{\psi_{0k}=\left\langle \psi_{0}|\psi_{k}\right\rangle \psi_{k}}.
\]
 We can now sum these terms and substitute in the expansion above $s_{z}\rightarrow s_{z}\left(m,k\right)$
to obtain for a cylindrical inclusion

\begin{equation}
\psi=\psi_{0}+\frac{4\pi}{\epsilon_{2}}\sum_{m}\int\frac{s_{z,km}^{2}}{s_{z}(m,k)-s_{z,km}}\left|\psi_{km}\right\rangle \nabla\psi_{km}^{*}\left(\mathbf{r}_{0}\right)\cdot\mathbf{p}dk.\mathbf{\ }\ .
\end{equation}
Note that the previous expansion for the electric potential with a uniform inclusion permittivity is satisfied for each $k$ component, which implies that one can vary $\epsilon$ as a function of $k$ in the expansion.

Finally, we analyze the response of a crystal inclusion. In the case
of a helical crystal the Fourier expansion is along a helical orbit
and the ``DC'' components have constant potential along this orbit.
We thus have coupling between modes of the types \cite{18} 
$\ensuremath{\left(m',k'\right)\rightarrow\left(m^{'}+pm,k^{'}+pmk_{z}\right)}\,\,$ and $\ensuremath{\left(m',k'\right)\rightarrow\left(m^{'},k^{'}+pnk_{z}\right)},$
 where $p$ is an integer number and $n$ is the number of units per
helical round. We will show next that for $\ensuremath{\rho_{0}-\rho_{2}>a/n},\,\,\,\rho_{0}-\rho_{2}>a/2$
the second and first types of coupling are negligible, respectively. We therefore conclude that for $\rho_{0}-\rho_{2}>a/2$
only the $m=1$ mode is important and write 
\begin{gather}
\psi\left(\boldsymbol{r},\rho_{0}>a/2\right)\approx\psi_{0}\left(\boldsymbol{r}\right)+\frac{4\pi}{\epsilon_{2}}\times\nonumber\\
\int\frac{s_{z,km=1}^{2}}{s_{z}(m=1,k)-s_{z,km=1}}\left|\psi_{km}\right\rangle \nabla\psi_{km}^{*}\left(\mathbf{r}_{0}\right)\cdot\mathbf{p}dk.
\end{gather}

We can substitute the eigenpermittivities and the physical permittivity,
to get $\ensuremath{s_{z}\left(m=1,k\right),s_{k,m=1}},$ respectively,
and obtain an expansion for $\ensuremath{\psi\left(\mathbf{r}\right)}.$
The calculation of the eigenpermittivities can be performed using the boundary conditions and the physical permittivity can be measured in some cases
or calculated by substituting $\ensuremath{\omega\left(k\right)}$ in $\ensuremath{\omega_{T}}$
in the expression for $\epsilon.$  $\omega\left(k\right)$
is calculated in the main text from the EOM and can also be calculated
when anharmonic terms are incorporated.

Since a strong response is expected at $m=1,k=k_{z},$ a dipole that emits at a range of
spatial frequencies will interact more dominantly with this mode.
In this region the dominant term in the expansion is
\[
\frac{4\pi}{\epsilon_{2}}\frac{s_{k_{z}m=1}^{2}}{s_{z}\left(m=1,k_{z}\right)-s_{k_{z}m=1}}\left|\psi_{k_{z}m=1}\right\rangle \nabla\psi_{k_{z}m=1}^{*}\left(\mathbf{r}_{0}\right)\cdot\mathbf{p},
\]
in addition to $\ensuremath{\psi_{0}},$ where $k_{z}=\frac{2\pi}{a}$,
$a$ is the helical-orbit axial periodicity.


\subsection*{A.1.2 The form of the eigenfunctions}

Since $\boldsymbol{E}_{\mathrm{inc}}$/$\psi_{0}$ component with
a given $k$ results in a contribution of an eigenfunction with the
same $k$ in the expansion, the eigenfunctions that account for the
field scattering due to synchronous vibrations are 
\[
\psi_{m}=e^{im\left(\phi-k_{z}z\right)}\left\{ \begin{array}{cc}
A_{1m}K_{m}\left(mk_{z}\rho\right) & \rho>\rho_{2}\\
A_{2m}I_{m}+A_{3m}K_{m} & \rho_{1}<\rho<\rho_{2}\\
A_{4m}I_{m}\left(mk_{z}\rho\right) & \rho<\rho_{1}
\end{array}\right.,
\]
where $\phi,z,\rho$ are cylindrical-coordinates variables, $I_{m},K_{m}$
are the modified Bessel functions, $\rho_{1},\rho_{2}$ are the internal
and external inclusion radii, $k_{z}=2\pi/a,$ and $a$ is the helical-orbit
axial period. Upon a continuous translation along the helical orbit,
$\psi_{m}$ remains constant and therefore corresponds to an eigenvalue
1. We can similarly take the directional derivative in the direction
of the helical orbit and obtain 
\begin{gather}
\nabla_{v}\psi_{m}=v\cdot\nabla\psi_{m}\nonumber\\
=-\frac{i}{\sqrt{(\rho k_{z})^{2}+1}}\left(\rho k_{z},1\right)\cdot\left(m/\rho,-mk_{z}\right)e^{im\left(\phi-k_{z}z\right)}=0,\nonumber
\end{gather}
as expected. This means that $\hat{R}\psi_{n}=\psi_{n},$ where $\hat{R}$
is the continuous-translation operator. 

\subsection*{A.1.3 Scaling of the eigenfunctions}

We analyze the scaling of $\psi_{m}$ for small and large $\rho$s.
We start with the first $m=0$ mode
\[
K_{m}\left(x\rightarrow0\right)\rightarrow\left\{ \begin{array}{cc}
-\left[\ln\left(\frac{x}{2}\right)+0.5772\right] & m=0\\
\frac{\Gamma\left(m\right)}{2}\left(\frac{2}{x}\right)^{m} & m\neq0
\end{array}\right.
\]
Since for $m=0,x=0$ and we expect a finite potential, this mode is
associated in all regions with $I_{m=0}\left(x\right)$ and is constant
everywhere (and therefore can be omitted). This mode can be treated
in the full Maxwell-equation analysis and was shown to scale as
$\sqrt{1/\rho}$ \cite{bergmancylinder2008}. We proceed to the $m\geq1$ modes at $\rho\gg a$
and obtain
\[
K_{m\geq1}\left(mk_z\rho\gg a\right)\rightarrow\frac{1}{\sqrt{2mk_{z}}}\sqrt{\frac{\pi}{\rho}}e^{-mk_{z}\rho},
\]
with a typical interaction distance on the order of $\ensuremath{a/m}.$
This determines the range in which a dipole interacts with each mode. When $k$ and $m$ are large this approximation holds and one can show that $\lim_{k,m\rightarrow\infty}\frac{\left|\nabla\psi_{m,k\mu}\left(\boldsymbol{r}_{0}\right)\right|^{2}}{\tilde{\left\langle \psi_{m}|\psi_{m}\right\rangle }}\propto mk\epsilon_{1k}e^{-2km\rho_{0}}.$ Taking into account that $\mathrm{Im}\left(\epsilon_{1}\left(k\right)\right)>0,$  $\frac{s_{m}^{2}\mathrm{Im}\left(s^{*}\right)}{\left(\mathrm{Re}\left(s\left(k'\right)\right)-s_{m}\left(k'\right)\right)^{2}+\left(\mathrm{Im}\left(s\right)\right)^{2}}$ is  bounded since even at the limit $s_m\rightarrow\infty$ it equals 1, and that $\epsilon_{1k}$ should converge when $k\rightarrow\infty$ (see Appendix in Ref. \cite{farhi2020coupling}), the integral over $k'$ and sum over $m$ in Eq. (10) are ensured to converge. Clearly, the larger $\rho_{0}$ is, the faster it converges. 

The scaling of the helical modes inside the structure close to the origin is
\begin{gather}
I_{m}\left(x\rightarrow0\right)\rightarrow\frac{1}{\Gamma\left(m+1\right)}\left(\frac{x}{2}\right)^{m},\nonumber\\
I_{m=0}\left(mk_{z}\rho\rightarrow0\right)\rightarrow\frac{1}{\Gamma\left(m+1\right)}\left(\frac{mk_{z}\rho}{2}\right)^{m}\nonumber\\
\propto m^{m}\left(\frac{k_{z}\rho}{2}\right)^{m},\,\,\,\Gamma\left(m+1\right)=m!.\nonumber
\end{gather}
\vspace{5mm} 
\subsection*{A.1.4 Calculating the radial argument inside the inclusion}
In a crystal one can express the effective permittivity as $\epsilon=\epsilon\left(\omega,k\right),$
which relates the response at a given $k$ to an excitation at the
same $k.$ In the case of a microtubule (MT), this form of $\epsilon\left(\omega,k\right)$
is justified because the period length $a$ is 8nm and therefore, $(\lambda_{0}/a)^{2}\gg1$
where $\lambda_{0}=c/\omega$ is the vacuum wavelength
\cite{agranovich2013crystal}. Note that in the derivation in Ref. \cite{agranovich2013crystal} it is assumed
that inside the inclusion $\rho_{\mathrm{ext}}\left(\omega\right)=0,\boldsymbol{J}_{\mathrm{ext}}\left(\omega\right)=0,$
which is satisfied in our case since the charges on the tubulin and
tubulin dimers oscillate only as a response to an external excitation
and can therefore be defined as polarization. Also, eigenstates are
defined for a system without a source. Another argument is that
for sources at distances larger than the typical interaction distance
of the $m=2$ mode, the inclusion is approximately not affected by the $m>1$ modes.

To represent axial vibrations, we assume an anisotropic inclusion
with an axial permittivity $\epsilon_{z}$ and radial and azimuthal
permittivitties $\epsilon_{2},$ equal to the host-medium permittivity,
where we omit $k$ for brevity. Note that the eigenpermittivities
in the quasistatic regime do not depend on $\omega.$ We now solve
Laplace\textquoteright s equation in cylindrical coordinates inside
the anisotropic inclusion. This will allow us to find the argument
of the functions $I_{m},K_{m}$ for $\rho_{1}<\rho<\rho_{2}$ and
calculate the eigenpermittivities. Substituting
the form of $\psi_{m}$ we write Laplace\textquoteright s equation
inside the helical structure
\begin{gather}
\nabla\overleftrightarrow{\epsilon}\nabla\psi_{m}=0,\nonumber\\
\epsilon_{2}\frac{1}{\rho}\frac{\partial}{\partial\rho}\left(\rho\frac{\partial\psi_{m}}{\partial\rho}\right)-\epsilon_{2}m^{2}\frac{1}{\rho^{2}}\psi_{m}-k_{z}^{2}m^{2}\epsilon_{zm}\psi_{m}=0.
\end{gather}
We change variables 
\[
x\equiv km\sqrt{\epsilon_{z}/\epsilon_{2}}\rho,\,\,\,\frac{\partial}{\partial\rho}=\frac{\partial}{\partial x}\frac{\partial x}{\partial\rho}=\frac{\partial}{\partial x}k_{z}m\sqrt{\epsilon_{zm}/\epsilon_{2}},
\]
and write 
\begin{widetext}
\begin{gather}
\frac{1}{x}k_{z}^{2}m^{2}\epsilon_{zm}\frac{\partial}{\partial x}\left(x\frac{\partial\psi_{m}}{\partial x}\right)-m^{2}\frac{\left(k_{z}m\sqrt{\epsilon_{zm}}\right)^{2}}{x^{2}}\psi_{m}-k_{z}^{2}m^{2}\epsilon_{zm}\psi_{m}=0,\nonumber \\
\frac{1}{x}\frac{\partial}{\partial x}\left(x\frac{\partial\psi_{m}}{\partial x}\right)-\left(\frac{m^{2}}{x^{2}}\psi_{m}+1\right)\psi_{m}=0.
\end{gather}
Thus we get
\[
\psi_{m}=e^{im\left(\phi-k_{z}z\right)}\left\{ \begin{array}{cc}
A_{1m}K_{m}\left(mk_{z}\rho\right) & \rho>\rho_{2}\\
A_{2m}I_{m}\left(mk_{z}\sqrt{\frac{\epsilon_{zm}}{\epsilon_{2}}}\rho\right)+A_{3m}K_{m} & \left(mk_{z}\sqrt{\frac{\epsilon_{zm}}{\epsilon_{2}}}\rho\right)\rho_{1}<\rho<\rho_{2}\\
A_{4m}I_{m}\left(mk_{z}\rho\right) & \rho<\rho_{1}
\end{array}\right.,
\]
\end{widetext}
which needs to be multiplied by additional factors to obtain the contribution
in the expansion of the potential of a point charge as we showed in
the previous subsection. Note that when calculating the total response as in Eqs. (2), (5), and (10) one has to sum over $m$ and integrate over $k$ for any relation between $k$ and $m.$
\subsection*{A.2 Isotropic cylindrical shell} 
\subsubsection*{A.2.1 Calculating the eigenpermittivities}
We express the eigenvalue equation and the relations between the coefficients of the eigenfunctions of an isotropic cylindrical shell
\[
\psi_{m,k}=e^{im\phi+kz}\left\{ \begin{array}{cc}
A K_{m}\left(k\rho\right) & \rho>\rho_{2}\\
B_1 I_{m}\left(k\rho\right)+B_2 K_{m}\left(k\rho\right) & \rho_{1}<\rho<\rho_{2}\\
C_1 I_{m}\left(k\rho\right) & \rho<\rho_{1}
\end{array}\right.,
\]
where $B_{1}$ is treated as known (cancels out in the expansion).
We first write the boundary conditions
\begin{gather}
Aa=B_{1}b_{11}+B_{2}b_{21},\nonumber\\
B_{1}b_{12}+B_{2}b_{22}=C_{1}c,\nonumber\\
A\epsilon_{2}a_{d}=\epsilon_{1}\left(B_{1}b_{11d}+B_{2}b_{21d}\right),\nonumber\\
\epsilon_{1}\left(B_{1}b_{12d}+B_{2}b_{22d}\right)=C_{1}\epsilon_{2}c_{d},\nonumber
\end{gather}
where
\begin{gather}
a=I_{m}\left(k\rho_{1}\right)\text{,}\,b_{11}^{\pm}=I_{m}\left(k\rho_{1}\right)\text{,}\,b_{12}^{\pm}=I_{m}\left(k\rho_{2}\right),\,\,b_{21}^{\pm}=K_{m}\left(k\rho_{1}\right),\nonumber\\
b_{22}^{\pm}=K_{m}\left(k\rho_{2}\right),\,\,c=K_{m}\left(k\rho_{2}\right),\,\,a_{d}=\epsilon_{2}\left(\frac{\partial I_{m}\left(k\rho\right)}{\partial\rho}\right)_{\,\rho=\rho_{1}},\nonumber\\
b_{11d}^{\pm}=\epsilon_{1}\left(\frac{\partial}{\partial\rho}I_{m}\left(k\rho\right)\right)_{\,\rho=\rho_{1}},b_{12d}^{\pm}=\left(\frac{\partial}{\partial\rho}I_{m}\left(k\rho\right)\right)_{\rho=\rho_{2}},\nonumber\\
b_{21d}^{\pm}=\epsilon_{1}\left(\frac{\partial}{\partial\rho}K_{m}\left(k\rho\right)\right)_{\rho=\rho_{1}}\text{,}\,b_{22d}^{\pm}=\epsilon_{1}\left(\frac{\partial}{\partial\rho}K_{m}\left(k\rho\right)\right)_{\rho=\rho_{2}},\nonumber\\
c_{d}=\epsilon_{2}\left(\frac{\partial K_{m}\left(k\rho\right)}{\partial\rho}\right)_{\,\rho=\rho_{2}}.\nonumber
\end{gather}
We write two relations between $B_{2}$ and $\epsilon_{1}$
\begin{equation}
\epsilon_{1}\left(B_{1}b_{12d}+B_{2}b_{22d}\right)=\epsilon_{2}\frac{B_{1}b_{12}+B_{2}b_{22}}{c}c_{d},
\end{equation}
\begin{equation}
\frac{B_{1}b_{11}+B_{2}b_{21}}{a}a_{d}\epsilon_{2}=\epsilon_{1}\left(B_{1}b_{11d}+B_{2}B_{21d}\right),
\end{equation}
and express $B_{2}$
\[
\epsilon_{1}B_{2}b_{22d}-\epsilon_{2}\frac{B_{2}b_{22}c_{d}}{c}=\epsilon_{2}\frac{B_{1}b_{12}}{c}c_{d}-\epsilon_{1}B_{1}b_{12d},
\]
\[
B_{2}\left(\epsilon_{1}b_{22d}-\epsilon_{2}\frac{b_{22}c_{d}}{c}\right)=B_{1}\left(\epsilon_{2}\frac{b_{12}}{c}c_{d}-\epsilon_{1}b_{12d}\right),
\]
\[
B_{2}=B_{1}\left(\frac{\epsilon_{2}\frac{b_{12}}{c}c_{d}-\epsilon_{1}b_{12d}}{\epsilon_{1}b_{22d}-\epsilon_{2}\frac{b_{22}c_{d}}{c}}\right).
\]
Substituting $B_2$ we obtain the quadratic eigenvalue equation
for $\epsilon_{1}$
\begin{widetext}
\[
0=\frac{\epsilon_{1}}{\epsilon_{2}}b_{22d}\left(b_{11d}\frac{\epsilon_{1}}{\epsilon_{2}}-b_{11}\frac{a_{d}}{a}\right)-\frac{b_{22}c_{d}}{c}\left(b_{11d}\frac{\epsilon_{1}}{\epsilon_{2}}-b_{11}\frac{a_{d}}{a}\right)+\frac{b_{12}}{c}c_{d}\left(\frac{\epsilon_{1}}{\epsilon_{2}}b_{21d}-b_{21}\frac{a_{d}}{a}\right)-\frac{\epsilon_{1}}{\epsilon_{2}}b_{12d}\left(\frac{\epsilon_{1}}{\epsilon_{2}}b_{21d}-b_{21}\frac{a_{d}}{a}\right).
\]
Finally, we express $A$ and $C_{1}$ 
\[
A=B_{1}\frac{b_{11}+\left(\frac{\epsilon_{2}\frac{b_{12}}{c}c_{d}-\epsilon_{1}b_{12d}}{\epsilon_{1}b_{22d}-\epsilon_{2}\frac{b_{22}c_{d}}{c}}\right)b_{21}}{a},\,\,\,C_{1}=B_{1}\frac{b_{12}+\left(\frac{\epsilon_{2}\frac{b_{12}}{c}c_{d}-\epsilon_{1}b_{12d}}{\epsilon_{1}b_{22d}-\epsilon_{2}\frac{b_{22}c_{d}}{c}}\right)b_{22}}{c},
\]
and obtain the two sets of solutions:
\begin{gather}
A=-\frac{B_{1}}{2a\left(ab_{22}b_{21d}c_{d}-b_{21}ca_{d}b_{22d}\right)}\left[-cb_{21}^{2}a_{d}b_{12d}+b_{11}b_{21}ca_{d}b_{22d}+ab_{22}b_{21}b_{11d}c_{d}+ab_{12}b_{21}b_{21d}c_{d}+-2ab_{11}b_{22}b_{21d}c_{d}\right.\nonumber\\
\pm \left.b_{21}\sqrt{\left(ca_{d}\left(b_{21}b_{12d}-b_{11}b_{22d}\right)+a\left(b_{12}b_{21d}-b_{22}b_{11d}\right)c_{d}\right){}^{2}-4a\left(b_{12}b_{21}-b_{11}b_{22}\right)ca_{d}\left(b_{12d}b_{21d}-b_{11d}b_{22d}\right)c_{d}}\right] ,\nonumber\\
B_{2}=\frac{B_{1}}{2b_{21}ca_{d}b_{22d}-2ab_{22}b_{21d}c_{d}}\left[-b_{21}ca_{d}b_{12d}-b_{11}ca_{d}b_{22d}+ab_{22}b_{11d}c_{d}+ab_{12}b_{21d}c_{d}\right.\nonumber\\
\pm \left.\sqrt{\left(ca_{d}\left(b_{21}b_{12d}-b_{11}b_{22d}\right)+a\left(b_{12}b_{21d}-b_{22}b_{11d}\right)c_{d}\right){}^{2}-4a\left(b_{12}b_{21}-b_{11}b_{22}\right)ca_{d}\left(b_{12d}b_{21d}-b_{11d}b_{22d}\right)c_{d}}\right],\nonumber\\
C=\frac{B_{1}}{2c\left(b_{21}ca_{d}b_{22d}-ab_{22}b_{21d}c_{d}\right)}\left[ab_{22}^{2}b_{11d}c_{d}-b_{21}b_{22}ca_{d}b_{12d}-b_{11}b_{22}ca_{d}b_{22d}-ab_{12}b_{22}b_{21d}c_{d}+2b_{12}b_{21}ca_{d}b_{22d}\right.\nonumber\\
\pm \left.b_{22}\sqrt{\left(ca_{d}\left(b_{21}b_{12d}-b_{11}b_{22d}\right)+a\left(b_{12}b_{21d}-b_{22}b_{11d}\right)c_{d}\right){}^{2}-4a\left(b_{12}b_{21}-b_{11}b_{22}\right)ca_{d}\left(b_{12d}b_{21d}-b_{11d}b_{22d}\right)c_{d}}\right],\nonumber\\
\epsilon_{1k}/\epsilon_{2}=-\frac{1}{2ac\left(b_{12d}b_{21d}-b_{11d}b_{22d}\right)}\left[-b_{21}ca_{d}b_{12d}+b_{11}ca_{d}b_{22d}+ab_{22}b_{11d}c_{d}-ab_{12}b_{21d}c_{d}+\right.\nonumber\\
\pm \left.\sqrt{\left(ca_{d}\left(b_{21}b_{12d}-b_{11}b_{22d}\right)+a\left(b_{12}b_{21d}-b_{22}b_{11d}\right)c_{d}\right){}^{2}-4a\left(b_{12}b_{21}-b_{11}b_{22}\right)ca_{d}\left(b_{12d}b_{21d}-b_{11d}b_{22d}\right)c_{d}}\right].\nonumber
\end{gather}
\end{widetext}

\subsection*{3 Including anharmonicity}

Here we calculated $\omega_k$ when anharmonicity in the axial forces between lateral units is included in the model without dispersion, where we set the amplitude $u_1 = 0.1a$ for simplicity. It can be seen that  $\mathrm{Im}(\omega_k)$ increases away from $k=k_z,$ which implies a stronger and delocalized response when $k\approx k_z,$ since the physical frequency is real. A complete account of this analysis as a function of the incoming field will be given elsewhere.  
\begin{figure}[h]
\begin{centering}
\includegraphics[width=8cm]{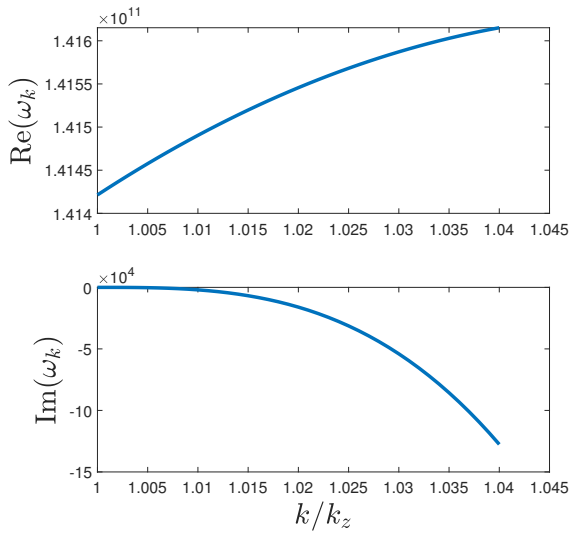}
\par\end{centering}
\caption{(a) Real part of $\omega_k$ and  (b) Imaginary part of $\omega_k$, where the coefficient of the anharmonic term is $k_5 =5k_3 k_z$ and the mode amplitude is $u_1 = 0.1a.$}
\end{figure}

%
%
 \FloatBarrier
\bibliographystyle{unsrt}

\end{document}